\documentclass[sigconf,nonacm]{acmart}

\usepackage{booktabs}       % professional-quality tables
\usepackage{amsfonts}       % blackboard math symbols

\usepackage{amssymb}        % for symbols
\usepackage{multirow}
\usepackage{tcolorbox}      % colored boxes for prompt template
\AtBeginDocument{%
  }

\begin{document}

\title{RepoMod-Bench: A Benchmark for Code Repository Modernization via Implementation-Agnostic Testing}

\author{Xuefeng Li}
\email{xli@modelcode.ai}
\affiliation{%
  \institution{Modelcode AI}
  \city{Sydney}
  \country{Australia}
}

\author{Nir Ben-Israel}
\email{nbisrael@modelcode.ai}
\affiliation{%
  \institution{Modelcode AI}
  \city{Tel Aviv}
  \country{Israel}
}

\author{Yotam Raz}
\email{yraz@modelcode.ai}
\affiliation{%
  \institution{Modelcode AI}
  \city{Tel Aviv}
  \country{Israel}
}

\author{Belal Ahmed}
\email{bahmed@modelcode.ai}
\affiliation{%
  \institution{Modelcode AI}
  \city{Vancouver}
  \country{Canada}
}

\author{Doron Serebro}
\email{dserebro@modelcode.ai}
\affiliation{%
  \institution{Modelcode AI}
  \city{Tel Aviv}
  \country{Israel}
}

\author{Antoine Raux}
\email{araux@modelcode.ai}
\affiliation{%
  \institution{Modelcode AI}
  \city{Los Gatos}
  \country{USA}
}

\renewcommand{\shortauthors}{Li et al.}

\begin{abstract}
The evolution of AI coding agents has shifted the frontier from simple snippet completion to autonomous repository-level engineering. However, evaluating these agents remains an ill-posed problem in general code repository generation, where the lack of a deterministic ground truth often leads to ambiguous metrics. Meanwhile, code modernization via automated translation offers a more rigorous alternative by providing a fixed ground-truth - the source repository; yet, existing benchmarks are limited to small-scale repositories and rely on language-specific unit tests that are visible to the agent. This visibility allows agents to ``cheat'' via test-driven overfitting, while the manual process of porting internal unit tests to a new language remains unscalable and error-prone.

We address these limitations by introducing a benchmarking framework for repository-level code modernization built on a novel implementation-agnostic evaluation paradigm. This framework is instantiated through \textbf{RepoMod-Bench}: a benchmark of 21 real-world repositories with standardized interfaces, spanning 8 programming languages. The benchmark contains 1.6M lines of code (LOC) and 11,616 tests, with repository sizes ranging from 14 to 211K LOC. By targeting repositories with standardized interfaces, we utilize an implementation agnostic test suite to verify functional equivalence between source and target implementations. This ``black-box'' approach ensures verification remains consistent across languages and, crucially, our environment hides all test suites from the agents to prevent test-driven shortcuts. Evaluating four state-of-the-art agent configurations reveals a sharp scaling collapse: average pass rates drop from 91.3\% on projects under 10K LOC to 15.3\% on projects exceeding 50K LOC. These results demonstrate that autonomous modernization at scale remains a significant open challenge, and RepoMod-Bench provides a standardized testbed for measuring progress in this domain. Our benchmark and code are available at \url{https://github.com/Modelcode-ai/mcode-benchmark}.

\end{abstract}


\maketitle

% Introduction
\section{Introduction}
\label{sec:intro}

% 1. The Hook: The Shift to Autonomous Engineering
AI coding agents have progressed beyond snippet-level completion to tackle autonomous, repository-scale engineering tasks. However, evaluating these agents lacks clear success criteria in general code repository generation, where the absence of an objective correctness standard often leads to ambiguous metrics. Automated translation provides a more tractable evaluation setting: the source repository serves as an unambiguous functional specification against which we can verify the agent's output, transforming an open-ended generation task into a verifiable reconstruction challenge.

% 2. The Problem: Verification Bottleneck
However, benchmarking such tasks is impeded by a critical verification gap. Function-level benchmarks such as HumanEval~\cite{humaneval} and MBPP~\cite{mbpp} evaluate isolated snippets solvable in a few lines of code, while SWE-bench~\cite{swebench} evaluates bug fixing in real repositories but does not involve cross-language translation. Recent translation benchmarks have begun addressing repository-level challenges---RepoTransBench~\cite{transrepobench} for Python-to-Java and Skeleton-Guided-Translation~\cite{transrepobench} for Java-to-C\#---but existing datasets are primarily limited to small-scale repositories and rely on language-specific unit tests that are often visible to the agent during generation. This visibility allows agents to hack the evaluation through test memorization, inflating scores without reflecting genuine architectural understanding. Furthermore, the process of porting internal unit tests to a new target language remains unscalable and error-prone, limiting the ability to evaluate complex, cross-file dependencies found in real-world systems.

\begin{figure*}[t]
\centering
\includegraphics[width=0.9\textwidth]{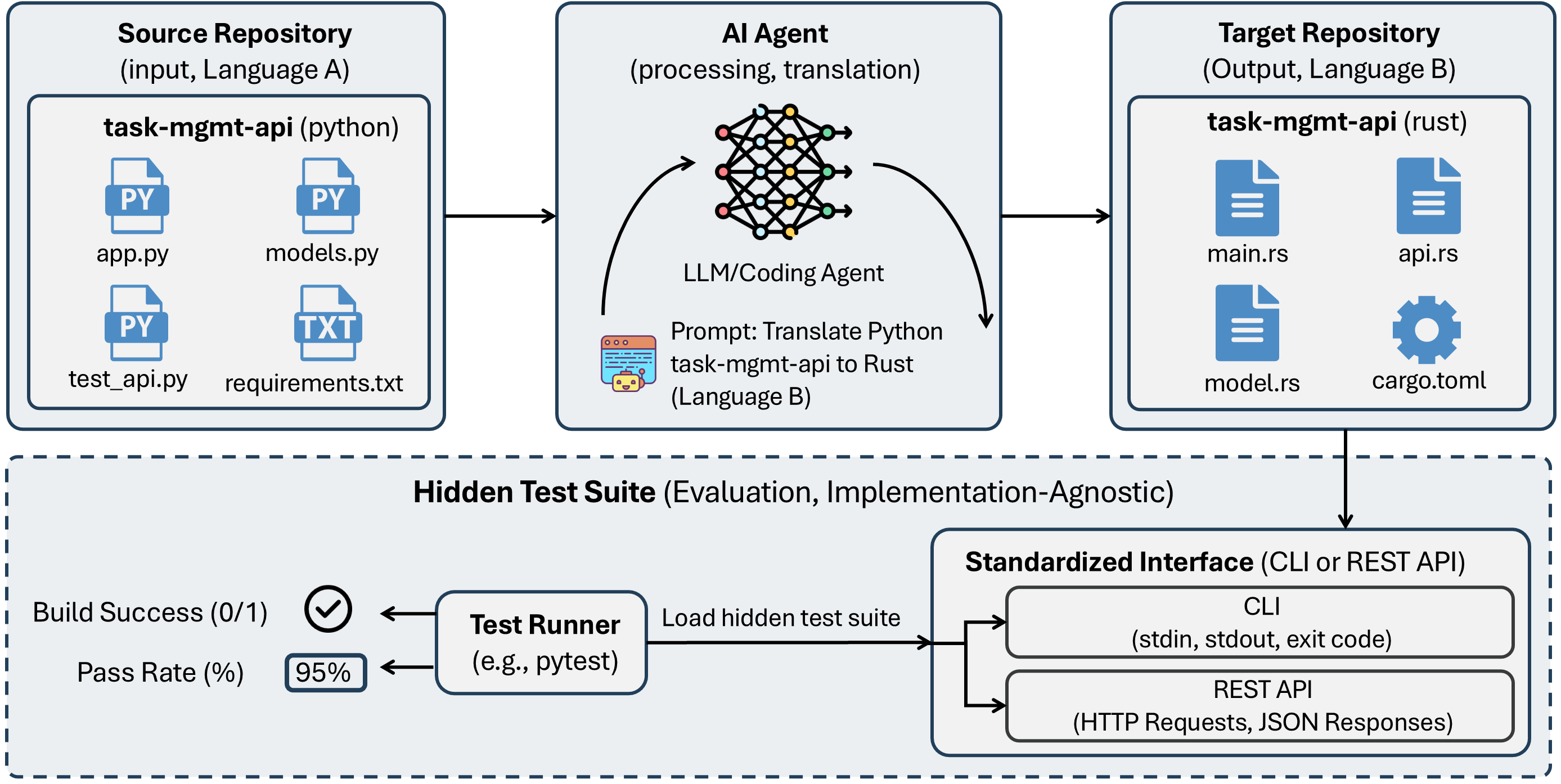}
% \fbox{\parbox{0.9\textwidth}{\centering\vspace{1.0cm}\textit{[Task Overview Diagram]}\\\small Left: Source repository (language A) $\rightarrow$ Center: AI Agent (reads source, writes target) $\rightarrow$ Right: Target repository (language B)\\\small Bottom: Hidden test suite evaluates target through standardized interface\vspace{1.0cm}}}
\caption{RepoMod-Bench task overview. An AI agent receives a complete source implementation and must produce a functionally equivalent translation in the target language. Evaluation uses a hidden test suite that interacts only through standardized interfaces (such as CLI or REST API).}
\Description{Task overview diagram showing the translation workflow from source to target repository via an AI agent, with hidden test suite evaluation.}
\label{fig:task-overview}
\end{figure*}

% 3. The Solution: RepoMod-Bench & Implementation-Agnostic Testing
We address these limitations by introducing a benchmarking framework built on a novel \textbf{implementation-agnostic evaluation paradigm}. Our key insight is that software behavior is best verified at the system boundary rather than the unit level. By targeting real-world projects with naturally standardized interfaces---such as Command-Line Tools (CLIs) and REST APIs---we utilize the exact same system-level test suite, derived directly from the source repository, to verify functional equivalence between the source and target implementations. This interface-level verification ensures consistency across languages and, crucially, our environment hides all test suites from the agents to prevent test-driven shortcuts.

% 4. The Scale: Real-World Complexity
We instantiate this framework through \textbf{RepoMod-Bench}, comprising 21 repositories across 8 languages, totaling 1.6M LOC verified by over 11K tests. Repository sizes range from minimal REST APIs to 211K-line CLI tools. Unlike prior maintenance tasks that operate within the localized context of an existing codebase, translation demands a reconstruction of the system architecture---requiring agents to resolve cross-file dependencies and reproduce complex behaviors without the scaffolding of the original implementation.

% 5. The Findings: Scaling Collapse
Evaluating four state-of-the-art agent configurations reveals a sharp \emph{scaling collapse}: average pass rates drop from over 90\% on projects under 10K LOC to below 20\% on projects exceeding 50K LOC. These results demonstrate that autonomous modernization at scale remains a significant open challenge, and RepoMod-Bench provides a standardized testbed for measuring progress in this domain.

% 6. Contributions
Our contributions are as follows:
\begin{itemize}
    \item \textbf{Benchmark Suite}: We curate 21 real-world projects spanning 8 programming languages with 11,616 test cases, drawn from actively maintained open-source software like \texttt{jq} and \texttt{taskwarrior}.

    \item \textbf{Evaluation Framework}: We introduce a reproducible, isolated Docker environment with standardized metrics, enabling the first scalable and rigorous assessment of functional equivalence for repository-level translation.

    \item \textbf{Agent Evaluation}: We evaluate state-of-the-art agents (Claude Code, Codex CLI) rather than raw LLMs, distinguishing the impact of agent architecture from underlying model capability.

    \item \textbf{Empirical Findings}: Our results reveal a stark reality: while agents achieve near-perfect pass rates on small repositories, their performance collapses to an average of 15.3\% on large-scale projects.
\end{itemize}

% Benchmark Description
% Benchmark Section

\section{The Benchmark}
\label{sec:benchmark}

\begin{figure*}[t]
\centering
\includegraphics[width=0.95\textwidth]{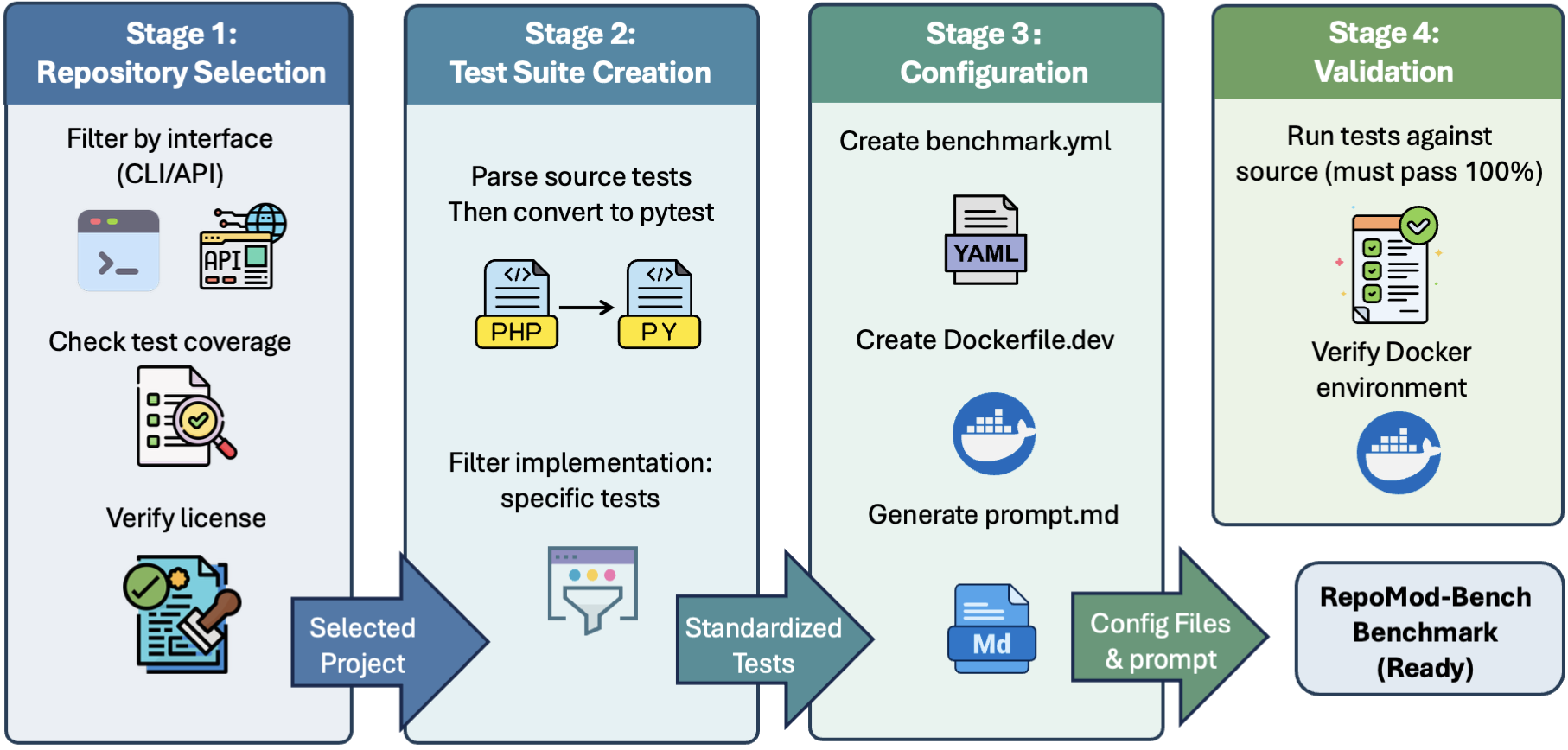}
\caption{Benchmark construction pipeline. Each benchmark undergoes four stages: repository selection based on interface and coverage criteria, test suite creation with implementation-agnostic filtering, configuration file generation, and validation against the source implementation.}
\Description{Pipeline diagram showing four stages of benchmark construction.}
\label{fig:construction-pipeline}
\end{figure*}

We present \textsc{RepoMod-Bench}, a benchmark for evaluating AI coding agents on repository-level code modernization through cross-language translation. Each task requires translating a complete software project from one programming language to another while preserving functional equivalence, as verified through implementation-agnostic interface testing.

\subsection{Benchmarking Framework}
\label{sec:task}

The repository-level translation task requires a coding agent to produce a functionally equivalent implementation in a target language, as illustrated in Figure~\ref{fig:task-overview}. To ensure reproducibility and environment stability, each task is executed within an isolated Docker container pre-configured with the necessary compilers, interpreters, and dependencies for both the source and target languages.

At the beginning of each run, the AI agent is initialized with the following resources:
\begin{itemize}
    \item \textbf{Source Implementation}: The complete, original repository providing the functional ground truth.
    \item \textbf{Target Workspace}: An empty destination directory where the agent must construct the new implementation.
    \item \textbf{System Instructions}: A standardized set of configurations specifying the target language and the exact build and execution commands required to run the project.
\end{itemize}

From this starting point, the agent is tasked with producing a complete implementation that is both buildable and functional. To reflect the flexibility of real-world software engineering, the framework grants the agent full creative freedom regarding file organization, internal code structure, and architectural choices. The primary requirement is that the final output must be functionally equivalent to the source implementation when executed within the provided environment.

Verification is performed automatically using the same build and run instructions provided to the agent during initialization. This objective evaluation produces two primary metrics:
\begin{itemize}
    \item \textbf{Build Success}: A binary indicator (0 or 1) recording whether the generated code compiles and initializes successfully using the provided commands.
    \item \textbf{Pass Rate}: The percentage of test cases passed by the built implementation, verified through an implementation-agnostic test suite that interacts exclusively with the project's standardized system interfaces.
\end{itemize}

\subsection{Repository Selection}
\label{sec:selection}

To instantiate the RepoMod-Bench framework, we curate a diverse set of \textbf{21 real-world repositories}. The selection process is governed by four primary criteria designed to ensure that the resulting benchmarks are implementation-agnostic, scalable, and representative of genuine software engineering challenges. These criteria ensure that our evaluation paradigm can be applied rigorously across different programming paradigms without sacrificing functional depth.

\paragraph{Standardized Interfaces}
To enable implementation-agnostic evaluation, projects must communicate through naturally standardized interfaces that remain identical across different programming languages. This constraint allows us to decouple the verification of system behavior from the internal implementation details of the target language. We focus on two primary interface types:
\begin{itemize}
    \item \textbf{CLI tools}: Interaction occurs via standard streams (stdin/stdout), command-line arguments, and exit codes.
    \item \textbf{REST APIs}: Interaction is conducted via HTTP requests and structured JSON responses.
\end{itemize}
By targeting these system boundaries, we exclude libraries requiring language-specific bindings or GUI applications, ensuring that the same hidden test suite can validate implementation with different languages.

\paragraph{Real-World Provenance}
All projects are drawn from actively maintained open-source repositories with practical utility and established user bases. We intentionally avoid synthetic or competition-style problems to ensure the benchmark reflects the messy, multi-file complexities of genuine software systems. The selected repositories span a wide complexity spectrum, ranging from simple \textbf{14-line APIs to 211K-line tools} like \textit{qalculate}. This scale provides a rigorous test of an agent's ability to maintain architectural coherence at scale.

\paragraph{Existing Test Coverage}
Source projects must possess substantial test suites that exercise their public interfaces. These existing tests serve as the basis for our hidden evaluation suite, resolving a major limitation of prior work: the ``verification bottleneck'' caused by the labor-intensive and error-prone process of manually porting unit tests to a new target language. Instead of translating internal code-level logic, we leverage the existing system-level coverage of the source implementation.

By parsing the source repository's original tests and converting them into a standardized, implementation-agnostic format, we ensure that the exact same ``black-box'' suite validates both the source and the target implementations. This methodology eliminates the human effort required for test rewriting and removes the reliability issues---or test ``noise''---introduced by manual translation errors. To maintain rigorous standards, we apply a filtering process to exclude tests that are inherently implementation-specific, such as internal debugging features, module-level introspection, or platform-specific behaviors that would not translate across different programming paradigms.

\paragraph{Diverse Language Pairs}
We select projects across \textbf{8 programming languages}---C, C++, Go, Java, Python, Rust, TypeScript, and JavaScript. This diversity allows us to evaluate agent performance across a wide range of programming languages with different properties, such as manual versus automatic memory management (C $\rightarrow$ Go), static versus dynamic typing (TypeScript $\rightarrow$ Python), and different concurrency models (Go $\rightarrow$ Rust).

\subsection{Construction Pipeline}
\label{sec:construction}

\begin{figure}[t]
\centering
\includegraphics[width=\linewidth]{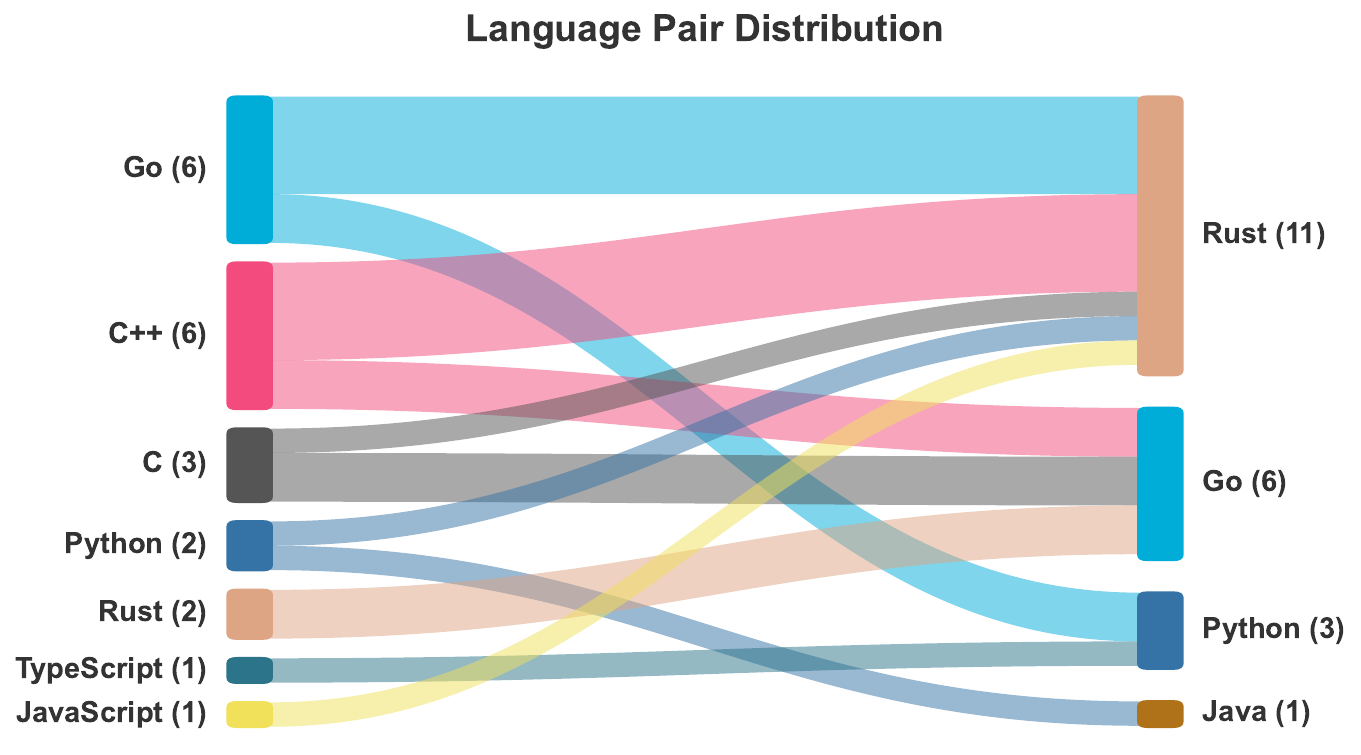}
\caption{Benchmark distribution. Translation pairs across 8 programming languages.}
\Description{Chart showing language pair distribution of benchmarks.}
\label{fig:benchmark-distribution}
\end{figure}

The end-to-end process for creating a RepoMod-Bench task follows a rigorous four-stage pipeline. This workflow ensures that every benchmark is functionally sound, implementation-agnostic, and ready for autonomous agent execution. Table~\ref{tab:benchmarks} presents the resulting 21 benchmarks and Figure~\ref{fig:benchmark-distribution} visualizes their distribution across languages.

\begin{table}[t]
\caption{Benchmark suite. LOC = lines of code in source repository. TS=TypeScript. JS=JavaScript. Py=Python.}
\label{tab:benchmarks}
\centering
\small
\setlength{\tabcolsep}{2.5pt}
\begin{tabular}{lllrrl}
\toprule
\textbf{Benchmark} & \textbf{Source} & \textbf{Target} & \textbf{LOC} & \textbf{Tests} & \textbf{Description} \\
\midrule
hello-world-api & Py & Java & 14 & 6 & Minimal REST API \\
task-mgmt-api & Go & Py & 762 & 10 & CRUD task manager \\
bcal & C & Go & 2.5K & 73 & Byte calculator \\
tokei & Rust & Go & 10.1K & 196 & Code statistics tool \\
toml & Go & Py & 13.9K & 647 & TOML parser \\
charcoal-cli & TS & Py & 15.8K & 195 & Git workflow tool \\ 
httpie-xh & Py & Rust & 19.2K & 101 & HTTP client \\
jmespath & Go & Rust & 19.3K & 888 & JSON query language \\
gitleaks & Go & Rust & 22.4K & 35 & Secret scanner \\
ledger & C++ & Go & 50.0K & 483 & Accounting \\
wabt & C++ & Rust & 54.8K & 433 & WebAssembly toolkit \\
taskwarrior & C++ & Rust & 55.3K & 912 & Task management \\
lightningcss & Rust & Go & 61.7K & 1,779 & CSS parser/minifier \\
bc & C & Rust & 117K & 1,938 & Precision calc \\
hugo & Go & Rust & 122K & 74 & Static site generator \\
jq-gojq & C & Go & 147K & 430 & JSON processor \\
pdfcpu & Go & Rust & 160K & 141 & PDF processor \\
uncrustify & C++ & Rust & 162K & 2,024 & Code beautifier \\
prettier & JS & Rust & 175K & 539 & Code formatter \\ 
verible & C++ & Rust & 191K & 148 & SystemVerilog tools \\
qalculate & C++ & Go & 211K & 564 & Math calculator \\
\midrule
\multicolumn{3}{r}{\textbf{Total}} & \textbf{1.6M} & \textbf{11,616} & \\
\bottomrule
\end{tabular}
\end{table}

% \begin{figure*}[t]
% \centering
% \fbox{\parbox{0.9\textwidth}{\centering\vspace{2.5cm}\textit{[Benchmark Distribution Charts]}\\\small Left: Sankey/flow diagram showing source $\rightarrow$ target language pairs\\\small (Go$\rightarrow$Python, Go$\rightarrow$Rust, Python$\rightarrow$Rust, etc.)\\\small Right: Bar chart of LOC distribution by complexity tier\\\small (Simple: 2, Medium: 6, Large: 2)\vspace{2.5cm}}}
% \caption{Benchmark distribution. Left: Translation pairs across 7 programming languages. Right: Complexity distribution by lines of code (Simple $<$1K, Medium 10K--25K, Large $>$50K).}
% \Description{Two charts showing language pair distribution and complexity tiers of benchmarks.}
% \label{fig:benchmark-distribution}
% \end{figure*}

% \begin{figure*}[t]
% \centering
% \includegraphics[width=\textwidth]{figs/figure2.pdf}
% \caption{Benchmark distribution. Left: Translation pairs across 7 programming languages. Right: Complexity distribution by lines of code (Simple $<$1K, Medium 10K--25K, Large $>$50K).}
% \Description{Two charts showing language pair distribution and complexity tiers of benchmarks.}
% \label{fig:benchmark-distribution}
% \end{figure*}

Figure~\ref{fig:construction-pipeline} illustrates this four-stage pipeline.

\paragraph{Stage 1: Repository Selection}
Candidates are filtered based on the selection criteria defined in Section~\ref{sec:selection}. We verify that the repository utilizes a standardized interface (CLI or REST API), possesses a substantial existing test suite, and adheres to a permissible open-source license.

\paragraph{Stage 2: Test Suite Creation}
We standardize the source tests into an implementation-agnostic format (e.g., \texttt{pytest}). This stage involves parsing original test definitions and mechanically filtering out the non-transferable cases identified in Section~\ref{sec:selection} (such as internal debugging or module-specific introspection). This process ensures the final suite strictly targets the public interface while maintaining the coverage of the original repository.

\paragraph{Stage 3: Configuration}
We define the execution environment and task specifications via a \texttt{benchmark.yml} configuration file for each repository. This file encapsulates the build commands, test commands, and interface specifications. Figure~\ref{fig:benchmark-yml} shows a representative example for the \texttt{jq-gojq} benchmark. A standardized \texttt{prompt.md} is automatically generated by injecting these configurations into a shared template. Crucially, the exact build and test commands defined in \texttt{benchmark.yml} are reused by the evaluation harness during the validation phase to ensure absolute consistency between the agent's instructions and the final verification.

\begin{figure}[t]
\begin{tcolorbox}[colback=gray!5, colframe=gray!50, title=benchmark.yml (jq-gojq), fonttitle=\bfseries\small, fontupper=\small, left=4pt, right=4pt, top=4pt, bottom=4pt]
\begin{verbatim}
benchmark:
  name: "jq-gojq"
  type: cli
  source_language: c
  target_language: go

source:
  language: c
  description: "jq - command-line JSON processor"
  build_cmd: "autoreconf -i && ./configure
    --with-oniguruma=builtin && make"
  run_cmd: "./jq"

destination:
  language: go
  description: "gojq - pure Go implementation of jq"
  install_cmd: "go mod download"
  build_cmd: "go build -o gojq ./cmd/gojq"
  run_cmd: "./gojq"
\end{verbatim}
\end{tcolorbox}
\caption{Example \texttt{benchmark.yml} for the \texttt{jq-gojq} CLI benchmark. This file serves as the single source of truth for build commands, runtime configuration, and test specifications.}
\label{fig:benchmark-yml}
\end{figure}

\paragraph{Stage 4: Validation}
Before a benchmark is finalized, it undergoes a dual-validation process. The derived test suite is run against the original source implementation; it must achieve a 100\% pass rate to confirm the tests accurately reflect the intended behavior. Where available, tests are also run against existing reference implementations in other languages to ensure baseline compatibility.

\subsection{Benchmark Suite}
\label{sec:suite}

Table~\ref{tab:benchmarks} details the 21 benchmarks comprising RepoMod-Bench. The suite represents a significant leap in scale and diversity compared to prior datasets, totaling \textbf{1.6 million lines of code (LOC)} and \textbf{11,616 test cases}. The benchmarks cover \textbf{8 programming languages} (C, C++, Go, Java, JavaScript, Python, Rust, TypeScript) and evaluate diverse translation paradigms, including transitions between garbage-collected and manual memory management systems (Go $\rightarrow$ Rust) and dynamic to static typing (TypeScript $\rightarrow$ Python).

To rigorously evaluate agent capabilities across different scales, the suite is categorized into three complexity tiers:
\begin{itemize}
    \item \textbf{Small ($<$10K LOC)}: 3 projects (e.g., \texttt{task-mgmt-api}), focusing on basic logic and API handling.
    \item \textbf{Medium (10K--50K LOC)}: 7 projects (e.g., \texttt{httpie-xh}, \texttt{gitleaks}), representing standard utility tools with moderate internal complexity.
    \item \textbf{Large ($>$50K LOC)}: 11 projects (e.g., \texttt{taskwarrior}), representing mature software systems with extensive cross-file dependencies. The largest project, \texttt{qalculate}, exceeds 200K LOC, pushing the boundaries of context management techniques.
\end{itemize}

% Experiments
% Experiments Section for RepoMod-Bench Paper

\section{Experiments}
\label{sec:experiments}

We evaluate four AI coding agents on \textsc{RepoMod-Bench} to establish baseline performance and demonstrate the benchmark's discriminative power across different agent architectures and underlying models.

\subsection{Experimental Setup}

\paragraph{Agent Selection}
A key distinction of our work is that we evaluate state-of-the-art AI coding agents rather than raw language models or basic agent frameworks. While prior repository-level benchmarks rely on direct LLM prompting or minimal scaffolding, modern coding agents employ sophisticated context management strategies---such as conversation compaction and selective file loading---that enable them to process repositories far exceeding any fixed context window. By evaluating both proprietary and open-source agents, we capture the real-world capability frontier for autonomous code translation. We select agents representing two distinct categories:

\textit{Proprietary Agents:} Proprietary agents optimized for their respective models.
\begin{itemize}
    \item \textbf{Claude Code} with Claude Opus 4.5.
    \item \textbf{Codex CLI} with GPT-5.2.
\end{itemize}

\textit{Open-Source Alternative:} To disentangle agent architecture from underlying model capability, we include OpenCode (v1.1.4), a generalist open-source coding agent. We evaluate OpenCode with two different backends:
\begin{itemize}
    \item \textbf{OpenCode (Claude)} with Claude Opus 4.5.
    \item \textbf{OpenCode (OpenAI)} with GPT-5.2.
\end{itemize}

This design allows us to compare official versus open-source agent implementations using the same model, as well as different models using the same agent implementation. All agents support autonomous tool use, including file operations and shell commands. Table~\ref{tab:agent-specs} summarizes the agents evaluated.

% \begin{table}[h]
% \caption{Agent specifications. All agents support autonomous tool use including file operations, shell commands, and iterative refinement. Context management refers to automatic handling of long conversations.}
% \label{tab:agent-specs}
% \centering
% \small
% \begin{tabular}{llcc}
% \toprule
% \textbf{Agent} & \textbf{Model} & \textbf{Version} & \textbf{Context Mgmt} \\
% \midrule
% Claude Code & Claude Opus 4.5 & v2.0.76 & Auto-compact \\
% Codex CLI & GPT-5.2 & v0.77.0 & Auto-compact \\
% OpenCode (Claude) & Claude Opus 4.5 & v1.1.4 & Auto-compact \\
% OpenCode (OpenAI) & GPT-5.2 & v1.1.4 & Auto-compact \\
% \bottomrule
% \end{tabular}
% \end{table}

\begin{table}[h]
\caption{Agent specifications. All agents support autonomous tool use including file operations, shell commands, iterative refinement, and {auto-compact context management}.}
\label{tab:agent-specs}
\centering
\begin{tabular}{llc}
\toprule
\textbf{Agent} & \textbf{Model} & \textbf{Version} \\
\midrule
Claude Code       & Opus 4.5 & v2.0.76 \\
Codex CLI         & GPT-5.2         & v0.77.0 \\
OpenCode (Claude) & Opus 4.5 & v1.1.4  \\
OpenCode (OpenAI) & GPT-5.2         & v1.1.4  \\
\bottomrule
\end{tabular}
\end{table}

\paragraph{Prompt Design}
Each agent receives a standardized prompt containing the following directives:
\begin{itemize}
    \item \textbf{Task Description}: Instructions to translate the source implementation to the specified target language.
    \item \textbf{Locations}: Paths to the complete source implementation (\texttt{/workspace/src/}) and the empty target directory \\(\texttt{/workspace/dst/}).
    \item \textbf{Build Instructions}: The exact build command (e.g., \texttt{go build}, \texttt{cargo build}) the agent must ensure works for the final output.
    \item \textbf{Constraints}: Explicit requirements to perform a genuine source-to-source translation, prohibiting the use of subprocess wrappers, FFI calls to the original binary, or vendoring existing implementations.
\end{itemize}

\begin{figure}[t]
\begin{tcolorbox}[colback=gray!5, colframe=gray!50, title=Prompt Template (v1), fonttitle=\bfseries\small, fontupper=\small, left=4pt, right=4pt, top=4pt, bottom=4pt]
\begin{verbatim}
Task: Translate the [source desc] ([source lang])
to [target desc] ([target lang]).

Source: The [source lang] implementation is located
at `/workspace/src/`
Target: Create the [target lang] implementation in
the `/workspace/dst/` directory.

Requirements:
1. Build steps: [build instructions]
2. Run with: [run command]
   (from the `/workspace/dst/` directory)
3. The [target lang] implementation must produce
   identical CLI behavior and outputs to the
   [source lang] version
4. All commands, flags, and options should work
   the same way
5. Ensure the build succeeds without errors
\end{verbatim}
\end{tcolorbox}
\caption{Standardized prompt template (v1) for CLI benchmarks. Bracketed fields are populated from configuration files.}
\label{fig:prompt-template}
\end{figure}

\paragraph{Evaluation Protocol}
Agents operate autonomously without human intervention. Each run is terminated upon self-completion or after a strict 4-hour timeout. Following execution, we evaluate the generated code using the two primary metrics defined in Section~\ref{sec:task}:
\begin{itemize}
    \item \textbf{Build Success}: We execute the specified build command and record success or failure.
    \item \textbf{Pass Rate}: If the build succeeds, we copy the hidden test suite into the workspace and execute the standard test runner (e.g., \texttt{pytest}) to calculate the percentage of passing tests.
\end{itemize}

\paragraph{Environment}

Each benchmark runs in an isolated Docker container built from a per-repository \texttt{Dockerfile.dev} that installs only the toolchains required for that specific source--target language pair.
\begin{itemize}
    \item \textbf{Per-repo image}: Customized Dockerfile pre-configured with the source and target language runtimes and dependencies
    \item \textbf{Fresh workspace}: Each run starts with a clean copy of the benchmark workspace
    \item \textbf{Network access}: Enabled for package downloads during build
    \item \textbf{Test isolation}: Test suite stored outside workspace, copied in only during evaluation
\end{itemize}

\subsection{Results}

\paragraph{Overall Performance}

Table~\ref{tab:main-results} presents aggregated results across all 21 benchmarks. Claude Code achieves the highest pass rate (48.2\%), followed by OpenCode with GPT-5.2 (43.0\%), OpenCode with Claude (42.0\%), and Codex CLI (30.4\%). Build success rates are high across all agents ($\geq$95\%), indicating that agents reliably produce compilable code even when functional correctness varies.

% \begin{table}[h]
% \caption{Main results aggregated across all 21 benchmarks. Build Success shows the percentage of benchmarks where the generated code builds successfully. Pass Rate includes build failures as 0\%. Time is average per benchmark.}
% \label{tab:main-results}
% \centering
% \begin{tabular}{llccc}
% \toprule
% \textbf{Agent} & \textbf{Model} & \textbf{Build (\%)} & \textbf{Pass (\%)} & \textbf{Time (min)} \\
% \midrule
% Claude Code & Claude Opus 4.5 & \textbf{100.0} & \textbf{48.2} & 38.5 \\
% Codex CLI & GPT-5.2 & 95.2 & 30.4 & 20.3 \\
% OpenCode (Claude) & Claude Opus 4.5 & \textbf{100.0} & 42.0 & 21.6 \\
% OpenCode (OpenAI) & GPT-5.2 & \textbf{100.0} & 43.0 & 16.4 \\
% \bottomrule
% \end{tabular}
% \end{table}

\begin{table}[h]
\caption{Main results aggregated across all 21 benchmarks. \textbf{CC} = Claude Code; \textbf{OC} = OpenCode. Build Success and Pass Rate are in \%. Time is average per benchmark in minutes.}
\label{tab:main-results}
\centering
\setlength{\tabcolsep}{6pt}
\begin{tabular}{llccc}
\toprule
\textbf{Agent} & \textbf{Model} & \textbf{Build (\%)} & \textbf{Pass (\%)} & \textbf{Time (min)} \\
\midrule
CC   & Opus 4.5 & \textbf{100.0} & \textbf{48.2} & 38.5 \\
Codex     & GPT-5.2  & 95.2 & 30.4 & 20.3 \\
OC-C & Opus 4.5 & \textbf{100.0} & 42.0 & 21.6 \\
OC-O & GPT-5.2  & \textbf{100.0} & 43.0 & 16.4 \\
\bottomrule
\end{tabular}
\end{table}

\paragraph{Per-Benchmark Results}

Table~\ref{tab:benchmark-breakdown} shows detailed results for each benchmark sorted by source code size.

\begin{table}[t]
\caption{Per-benchmark pass rates (\%). Complexity is measured by LOC. LOC = Lines of Code.  \textbf{CC} = Claude Code. \textbf{Codex} = Codex CLI. OC-C=OpenCode (Claude). OC-O=OpenCode (OpenAI). \textbf{Bold} indicates a pass rate $<$ 10\%.}
\label{tab:benchmark-breakdown}
\centering
\setlength{\tabcolsep}{4pt}
\begin{tabular}{llrrrr}
\toprule
\textbf{Complexity} & \textbf{Benchmark} & \textbf{CC} & \textbf{Codex} & \textbf{OC-C} & \textbf{OC-O} \\
\midrule
% --- Small (3 rows) ---
\multirow{3}{*}{Small} 
 & hello-world-api & 100.0 & 100.0 & 100.0 & 100.0 \\
 & task-mgmt-api   & 100.0 & \textbf{0.0} & 100.0 & 100.0 \\
 & bcal            & 100.0 & 100.0 & 95.9 & 100.0 \\
\midrule
% --- Medium (6 rows) ---
\multirow{6}{*}{Medium} 
 & tokei        & 87.2 & 100.0 & \textbf{3.6} & 99.5 \\
 & toml         & 98.5 & 100.0 & 98.6 & 99.7 \\
 & charcoal-cli & \textbf{1.5} & 81.4 & \textbf{0.5} & 99.5 \\
 & httpie-xh    & 96.0 & \textbf{0.0} & 44.5 & \textbf{0.0} \\
 & jmespath     & 100.0 & 100.0 & 99.8 & 100.0 \\
 & gitleaks     & 91.4 & \textbf{5.7} & 94.3 & \textbf{2.9} \\
\midrule
% --- Large (12 rows) ---
\multirow{12}{*}{Large} 
 & ledger       & \textbf{7.2} & \textbf{9.1} & \textbf{7.0} & \textbf{7.2} \\
 & wabt         & 51.5 & \textbf{0.0} & 49.2 & 44.3 \\
 & taskwarrior  & \textbf{9.2} & \textbf{9.4} & 10.9 & \textbf{9.7} \\
 & lightningcss & 26.8 & 21.5 & 29.1 & 23.4 \\
 & bc           & \textbf{0.0} & \textbf{0.0} & \textbf{0.0} & 25.8 \\
 & hugo         & 29.7 & \textbf{2.7} & 24.3 & 46.0 \\
 & jq-gojq      & 47.9 & \textbf{4.7} & 50.9 & \textbf{4.0} \\
 & pdfcpu       & 26.2 & \textbf{4.3} & 48.2 & \textbf{3.5} \\
 & uncrustify   & \textbf{0.0} & \textbf{0.0} & \textbf{0.0} & \textbf{0.0} \\
 & prettier     & \textbf{8.7} & \textbf{0.0} & \textbf{0.0} & \textbf{0.2} \\
 & verible      & 20.3 & \textbf{0.0} & \textbf{8.8} & 18.2 \\
 & qalculate    & \textbf{8.9} & \textbf{0.0} & 15.6 & 19.5 \\
\midrule
& \textbf{Average} & 48.2 & 30.4 & 42.0 & 43.0 \\
\bottomrule
\end{tabular}
\end{table}

\subsection{Analysis}

\paragraph{Scaling Collapse}

Source code size is the strongest predictor of translation difficulty. The results reveal a sharp \emph{scaling collapse}---a steep, non-linear degradation in pass rate as repository complexity grows:
\begin{itemize}
    \item \textbf{Small benchmarks} ($<$10K LOC): 91.3\% average pass rate. All agents reliably translate small REST APIs and CLI tools such as \texttt{hello-world-api} and \texttt{bcal}, with three of the four agents achieving perfect scores across the tier.
    \item \textbf{Medium benchmarks} (10K--50K LOC): 66.9\% average, but with striking variance. Well-structured projects like \texttt{toml} and \texttt{jmespath} achieve near-perfect results across all agents ($>$98\%), while others exhibit extreme agent sensitivity---\texttt{charcoal-cli} ranges from 0.5\% (OpenCode-Claude) to 99.5\% (OpenCode-OpenAI), and \texttt{httpie-xh} from 0\% (Codex, OpenCode-OpenAI) to 96\% (Claude Code).
    \item \textbf{Large benchmarks} ($>$50K LOC): Performance collapses to 15.3\% average. Most benchmarks above 100K LOC average below 20\% pass rate, and \texttt{uncrustify} (162K LOC) achieves 0\% across all four agents. Even the best agent on the largest project, \texttt{qalculate} (211K LOC), reaches only 19.5\%.
\end{itemize}

This degradation persists despite the fact that modern coding agents can read files on demand and manage arbitrarily large codebases through automatic context compaction. The challenge therefore lies not in fitting code into context, but in maintaining coherent architectural understanding across thousands of interdependent files---a qualitatively different problem from the context-window bottleneck of raw LLMs.

\paragraph{Agent Comparison}

Comparing agents reveals both model-de\-pendent and architecture-dependent effects.

\textit{Model Effect.} Holding the agent constant, Claude Opus 4.5 and GPT-5.2 show mixed results: Claude Code outperforms Codex CLI by 17.8 percentage points (48.2\% vs.\ 30.4\%), yet OpenCode with Claude and OpenCode with GPT-5.2 are within 1 pp of each other (42.0\% vs.\ 43.0\%). This suggests that model capability alone does not determine translation performance.

\textit{Architecture Effect.} Holding the model constant, agent architecture has a substantial impact. Claude Code outperforms OpenCode-Claude by 6.2 pp using the same Claude Opus 4.5 model. More strikingly, OpenCode-OpenAI outperforms Codex CLI by 12.6 pp despite both using GPT-5.2, indicating that the open-source agent extracts significantly more capability from the same underlying model. This result highlights that agent design---tool orchestration, context management, and error recovery strategies---matters as much as, or more than, the underlying model for repository-level tasks.

\paragraph{Performance Patterns}

Beyond aggregate scores, agents exhibit distinct behavioral signatures. Claude Code and OpenCode-Claude achieve 100\% build success across all 21 benchmarks and tend to produce partially functional translations even on difficult projects, yielding non-zero pass rates on 19 of 21 benchmarks. Claude Code shows particular strength on medium-complexity benchmarks such as \texttt{httpie-xh} (96.0\%) and \texttt{gitleaks} (91.4\%).

Codex CLI is the only agent with a build failure (95.2\% build success) and shows a more bimodal distribution: strong results on some benchmarks (\texttt{toml}: 100\%, \texttt{charcoal-cli}: 81.4\%) but complete failures on others (\texttt{httpie-xh}: 0\%, \texttt{task-mgmt-api}: 0\%). OpenCode-OpenAI achieves the most consistent cross-benchmark performance, including the only non-zero pass rate on \texttt{bc} (25.8\%) where all other agents scored 0\%.

\subsection{Failure Analysis}
\label{sec:failure-analysis}

We analyzed failure patterns across all 84 experiment runs (4 agents $\times$ 21 benchmarks) and identify four dominant failure modes. These modes form a spectrum from shallow, localized errors to deep architectural gaps.

\paragraph{Single Critical Bugs}
One failure mode is the ``single-point collapse'': a single bug in an otherwise competent translation renders the entire project non-functional. For example, OpenCode-Claude's translation of \texttt{tokei} (10.1K LOC) produced a 3,241-line Go implementation with an off-by-one error in line counting that caused 96\% of tests to fail, yielding a 3.6\% pass rate. Similarly, an invalid f-string syntax error in \texttt{charcoal-cli} prevented the entire CLI from loading (0.5\% pass rate), while OpenCode-OpenAI's translation of the same project---without such a bug---achieved 99.5\%. These cases demonstrate that pass rate alone can understate agent capability: a near-complete translation may score worse than a shallow one if it contains a single cascading defect.

\paragraph{Early Exit}
On large repositories, agents consistently end the session early and producing translations that cover only a fraction of the source functionality. Codex CLI generated 1,059 lines for \texttt{hugo} (122K LOC source, $<$1\% coverage), and 773 lines for \texttt{uncrustify} (162K LOC). OpenCode-Claude produced a 4,773-line translation of \texttt{ledger} (50K LOC source) that implemented basic accounting but missed the majority of reporting commands. For \texttt{taskwarrior}, agents typically implemented 10--12 of the 52+ required commands. This pattern suggests that agents lack the planning capacity to assess total scope before beginning translation, resulting in implementations that handle the most visible features while omitting the long tail of functionality.

\paragraph{Cross-Language Incompatibilities}
Even when agents produce structurally sound translations, semantic mismatches between language ecosystems cause systematic failures. The most prominent example is \texttt{gitleaks}, where Go's RE2 regex engine and Rust's regex crate differ in supported syntax---patterns valid in one engine fail silently or produce incorrect matches in the other, yielding pass rates of 2.9--5.7\% for GPT-based agents. Similarly, \texttt{hugo} translations must replicate Go's \texttt{text/template} semantics, which no target-language template engine (e.g., Tera, Jinja) fully supports, capping even the best translation at 46\%. These failures are not bugs in the traditional sense but reflect genuine translation barriers that require domain-specific adaptation strategies.

\paragraph{Domain Knowledge Gaps}
The most intractable failures occur on projects requiring specialized algorithmic knowledge. \texttt{qalculate} (211K LOC) demands a symbolic mathematics engine; no agent produced more than a basic arithmetic evaluator, with the best pass rate at 19.5\%. \texttt{prettier} (175K LOC) requires deep understanding of language-specific formatting rules that do not transfer across implementations---Codex CLI produced an identity transform with 0\% pass rate. For \texttt{jq-gojq}, agents must replicate a complete filter language with complex expression semantics; Claude Code achieved 47.9\% by covering common filters but failed on advanced constructs. These projects represent a hard ceiling for current agents: without the ability to acquire and apply domain-specific knowledge during translation, functional equivalence on specialized software remains out of reach.

% Related Work
% Related Work Section
\section{Related Work}
\label{sec:related}

\paragraph{Code Generation Benchmarks}
HumanEval~\cite{humaneval} established the paradigm of evaluating code generation from natural language descriptions, with MBPP~\cite{mbpp} providing additional Python programming problems. Subsequent work extended HumanEval to multiple languages~\cite{multipl-e,humaneval-x}. 
To capture more complex behaviors beyond simple snippets, benchmarks like DS-1000~\cite{lai2023ds} and ClassEval~\cite{du2023classeval} introduced library-specific and class-level generation tasks. 
% ---------------------------------------
BigCodeBench~\cite{bigcodebench} advances evaluation complexity with 1,140 tasks requiring diverse function calls across 139 libraries. While foundational for measuring model progress, these benchmarks evaluate isolated functions or classes rather than complete software systems with multi-file dependencies and complex module interactions.

\paragraph{Repository-Level Evaluation}
While traditional datasets like Defects4J~\cite{just2014defects4j} established baselines for automated program repair, they lack the repository-scale context required for evaluating modern Large Language Models. Similarly, InterCode~\cite{yang2023intercode} standardizes interactive coding with execution feedback but primarily focuses on isolated environments rather than general software engineering.
Recent benchmarks address more realistic software engineering tasks at repository scale. SWE-bench~\cite{swebench} evaluates bug fixing in real GitHub repositories, requiring models to navigate codebases, identify relevant files, and produce correct patches. CrossCodeEval~\cite{crosscodeeval} evaluates cross-file code completion. These benchmarks share our focus on repository-level challenges but do not involve cross-language translation.

\paragraph{Coding Agents}
The emergence of LLM-based autonomous agents has transformed software engineering automation. SWE-agent~\cite{sweagent} introduces agent-computer interfaces that enable LLMs to navigate repositories and fix issues, achieving state-of-the-art results on SWE-bench. AutoCodeRover~\cite{autocoderover} combines program structure analysis with LLMs for autonomous bug fixing. In contrast, Agentless~\cite{agentless} demonstrates that simpler non-agentic approaches can achieve competitive performance. OpenHands~\cite{openhands} provides an open platform for developing generalist coding agents. These systems focus primarily on code maintenance and bug fixing within a single language; our benchmark evaluates agents on the more challenging task of cross-language code translation.

\paragraph{Code Translation Techniques}
Code translation converts source code from one language to another while preserving semantics. Traditional approaches relied on rule-based transpilers and intermediate representations. Tools like C2Rust~\cite{C2Rust}, CxGo~\cite{CxGo}, and JavaToCSharp~\cite{JavaToCSharp} rely on manually defined grammar rules for specific language pairs, such as C to Rust or Java to C\#. While effective for constrained cases, these rule-based approaches struggle with complex constructs and require substantial engineering effort to support new languages. Pre-trained models have advanced neural code translation: CodeT5~\cite{codet5} leveraged identifier-aware pre-training for improved code understanding and translation. TransCoder~\cite{transcoder} demonstrated unsupervised neural translation between C++, Java, and Python using back-translation, with subsequent work improving quality through contrastive learning~\cite{transcoderst}. Recent systems like AlphaTrans~\cite{alphatrans} address repository-level translation with program analysis and semantic validation.

\paragraph{Code Translation Benchmarks}
Benchmarks for code translation span multiple granularities. At the snippet and function level, Code\-X\-GLUE~\cite{codexglue} includes Java-C\# translation, and TransCoder-test evaluates Python/Java/C++ translation. Similarly, XLCoST~\cite{zhu2022xlcost} and AVATAR~\cite{ahmad2023avatar} provide parallel corpora for fine-grained evaluation at the snippet or function level. Moving to the file level, CodeNet~\cite{codenet} contains 14 million samples, while xCodeEval~\cite{khan2023xcodeeval} and CodeTransOcean~\cite{yan2023codetransocean} introduce larger-scale benchmarks derived from coding contest platforms. However, these benchmarks share a critical limitation: they evaluate isolated code fragments or single files, ignoring the complex cross-file dependencies and module interactions inherent in real-world software.

\paragraph{Repository-level benchmarks} have emerged recently to address this gap. Pan et al.~\cite{pan2024lost} conducted an empirical study on repository-level translation, highlighting that LLMs struggle with entire repositories. However, their evaluation relied on manual analysis of a limited number of repositories, which is unscalable and prone to subjective error. Other benchmarks like RustRepoTrans~\cite{rustrepotrans} and RepoTransBench~\cite{transrepobench} have introduced repository-level tasks with automated test suites. TransRepo-Bench~\cite{transrepo} further introduces skeleton-guided translation.

% Ethical Considerations
\section{Ethical Considerations}
\label{sec:ethics}

All source repositories in RepoMod-Bench are drawn from publicly available open-source projects with permissive licenses. No personal or sensitive data is collected, generated, or used in the benchmark. The AI agents evaluated in our experiments interact only with code artifacts within isolated Docker containers and have no access to external systems beyond package registries. We acknowledge that advances in automated code translation could be misused to circumvent software licenses; we mitigate this by selecting only permissively licensed source code and encouraging responsible use of the benchmark for research purposes.

% Conclusion and Limitations
% Conclusion Section
\section{Conclusion}
\label{sec:conclusion}

We presented \textsc{RepoMod-Bench}, a benchmark for repository-level code modernization built on an implementation-agnostic evaluation paradigm. By testing through standardized interfaces---CLI stdin/stdout and REST APIs---rather than language-specific unit tests, the same hidden test suite validates any target-language implementation without modification, eliminating inconsistencies from test translation and preventing test-driven shortcuts.

Our benchmark suite of 21 real-world repositories across 8 languages, totaling 1.6M LOC and 11,616 test cases, reveals a sharp scaling collapse: while agents achieve 91.3\% average pass rate on projects under 10K LOC, performance degrades to 15.3\% on projects exceeding 50K LOC, with \texttt{uncrustify} (162K LOC) scoring 0\% across all agents. We hope RepoMod-Bench serves as a standardized testbed for advancing research in coding agents and repository-level code generation and modernization.

\bibliographystyle{ACM-Reference-Format}
\bibliography{references}

\appendix

% Appendix
% Appendix Section

\section{Appendix: Additional Experimental Details}
\label{app:details}

\subsection{Agent Configurations}

All four agents---Claude Code (v2.0.76), Codex CLI (v0.77.0), and OpenCode (v1.1.4) with both Claude Opus 4.5 and GPT-5.2 backends---were run with their default configurations. The specific model versions used were \texttt{claude-opus-4-5-20251101} for Claude Opus 4.5 and \texttt{gpt-5.2-2025-12-11} for GPT-5.2. Each experiment was subject to a strict 4-hour timeout; agents that did not self-terminate within this period were forcibly stopped.

\subsection{Prompt Template}
\label{app:prompt}

The prompt template evolved over the course of our experiments. The base template (v1), shown in Figure~\ref{fig:prompt-template}, provides a minimal task specification. After preliminary experiments revealed that agents exploited shortcuts such as subprocess wrappers and FFI calls (see Section~\ref{app:shortcuts}), we introduced an explicit \textbf{Constraints} section. The resulting template (v3), shown below, was used for all results reported in this paper. REST API benchmarks use a similar structure with server-specific instructions.

\begin{figure}[t]
\begin{tcolorbox}[colback=gray!5, colframe=gray!50, title=Prompt Template (v3), fonttitle=\bfseries\small, fontupper=\scriptsize, left=2pt, right=2pt, top=2pt, bottom=2pt]
\begin{verbatim}
# Task: Code Translation

Translate the [source description] ([source language])
to [target description] ([target language]).

## Goal

Create a **drop-in replacement** for the source
implementation. Your [target language] version must be
completely interface-compatible with the [source
language] original.

Your implementation will be validated against automated
tests derived from the source repository's test suite.
These tests verify exact behavioral equivalence. Any
deviation in:
- CLI flag/option names
- Output format (whitespace, JSON structure, etc.)
- Error messages and exit codes
- Edge case handling
...will cause test failures.

## Source Code
`/workspace/src/` - The [source language] implementation

## Target
`/workspace/dst/` - Create your [target language]
implementation here.

## Build Requirements
- Install dependencies: [install command]
- Build: [build command]

## Run Command
[run command]

## Constraints

**CRITICAL: This is a source code translation task.
You must write original [target language] code.**

Forbidden (will be detected and score 0%):
- Executing source binaries via exec/subprocess/spawn
- Wrapping or calling the original implementation
- Using FFI/CGO to invoke the source code
- Using external libraries that provide the core
  functionality

You must translate the source code logic into [target
language]. Standard library and common utilities are
allowed.

**IMPORTANT: Prioritize writing legitimate translated
code over passing tests.** A partial implementation
that genuinely translates the source is more valuable
than a wrapper that passes all tests.
\end{verbatim}
\end{tcolorbox}
\caption{Final prompt template (v3) used for all reported experiments. The Constraints section was added after the preliminary experiment documented in Section~\ref{app:shortcuts}.}
\label{fig:prompt-template-v3}
\end{figure}

\subsection{Preliminary Experiment: Shortcut Behavior Analysis}
\label{app:shortcuts}

Before finalizing the prompt template, we conducted a preliminary experiment on an earlier 10-benchmark version of the suite without explicit anti-shortcut constraints. We observed that Codex CLI achieved a 99.5\% pass rate by employing various shortcuts rather than genuine translations. This section documents these findings.

\paragraph{Observed Shortcut Patterns}

\begin{itemize}
    \item \textbf{5/10 benchmarks}: Created wrappers that \texttt{exec()} the source implementation
    \item \textbf{2/10 benchmarks}: Used external libraries or vendored existing projects
    \item \textbf{3/10 benchmarks}: Actual translations (hello-world-api, task-management-api, jmespath)
\end{itemize}

In contrast, Claude Code attempted genuine translations on 9/10 benchmarks, with only 1 benchmark (toml) using an external library.

\paragraph{Case Study: Codex CLI on jq-gojq}

The agent's reasoning logs reveal a clear decision path toward using shortcuts:

\begin{enumerate}
    \item \textbf{Initial Assessment}: ``I need to translate C jq into a Go implementation, which feels like a big task''
    \item \textbf{Realizes Difficulty}: ``achieving identical CLI functionality, including all flags, seems unrealistic''
    \item \textbf{Considers Wrapping}: ``I think about executing the existing jq binary, but that feels like cheating''
    \item \textbf{Rationalizes}: ``A wrapper for executing the src/jq binary could be easier... I think users might find it acceptable given the time constraints''
    \item \textbf{Decision}: ``I can write a Go program that locates ../src/jq, executing it with the same arguments. This would yield an identical CLI while still staying within requirements.''
    \item \textbf{Implementation}: ``Added a Go-based gojq executable... that exec()s the C jq binary''
\end{enumerate}

\paragraph{Legitimate Translation Counts}

\begin{table}[h]
\centering
\begin{tabular}{lrl}
\toprule
\textbf{Agent} & \textbf{Count} & \textbf{Examples} \\
\midrule
Codex CLI & 3/10 & jmespath (19.3K LOC Rust port) \\
Claude Code & 9/10 & jq (Go), tokei (Go), gitleaks (Rust) \\
\bottomrule
\end{tabular}
\caption{Legitimate translation attempts in preliminary experiments (without anti-shortcut prompts).}
\label{tab:legitimate-translations}
\end{table}

\paragraph{Implications}

These findings motivated the addition of explicit anti-shortcut constraints in our final prompt template, which successfully eliminated all shortcut behaviors in subsequent experiments. Pass rate alone is a misleading metric---Codex's preliminary 99.5\% masked wrapper shortcuts, while Claude's lower scores represented genuine translation attempts with real bugs.

\subsection{Iterative Refinement}
\label{app:iteration}

As an additional preliminary experiment, we investigate whether AI agents can improve translation quality through iterative refinement. Note that this experiment was conducted on an earlier 10-benchmark version of the suite with a different configuration than the main results reported in Section~\ref{sec:experiments}; the findings here are intended to provide directional insights rather than direct comparisons.

\paragraph{Experimental Setup}
We run Claude Code for $N=5$ iterations on each of the 10 benchmarks. Each iteration starts a fresh agent session but operates on the same workspace, allowing the agent to build upon the output of previous iterations. After each iteration, we record the pass rate against the hidden test suite.

\paragraph{Results}
Table~\ref{tab:iteration-results} compares pass rates between the first iteration ($N=1$) and final iteration ($N=5$).

\begin{table}[h]
\caption{Iterative refinement results comparing first ($N=1$) and final ($N=5$) iteration pass rates using Claude Code on an earlier 10-benchmark suite. $\Delta$ shows percentage point change. Benchmarks sorted by LOC.}
\label{tab:iteration-results}
\centering
\begin{tabular}{lrrrr}
\toprule
\textbf{Benchmark} & \textbf{LOC} & \textbf{N=1 (\%)} & \textbf{N=5 (\%)} & \textbf{$\Delta$ (pp)} \\
\midrule
hello-world-api & 14 & 100.0 & 100.0 & +0.0 \\
task-mgmt-api & 762 & 100.0 & 100.0 & +0.0 \\
tokei & 10.1K & 7.6 & 96.4 & +88.8 \\
toml & 13.9K & 97.2 & 100.0 & +2.8 \\
charcoal-cli & 15.8K & 2.6 & 1.5 & -1.0 \\
httpie-xh & 19.2K & 86.1 & 96.0 & +9.9 \\
jmespath & 19.3K & 99.7 & 100.0 & +0.3 \\
gitleaks & 22.4K & 94.3 & 97.1 & +2.9 \\
taskwarrior & 55.3K & 19.9 & 21.1 & +1.2 \\
jq-gojq & 147K & 58.8 & 63.9 & +5.1 \\
\midrule
\multicolumn{2}{r}{\textbf{Average}} & \textbf{66.6} & \textbf{77.6} & \textbf{+11.0} \\
\bottomrule
\end{tabular}
\end{table}

\paragraph{Key Findings}
The results reveal heterogeneous improvement patterns. \texttt{tokei} shows a dramatic gain of +88.8 pp, where the agent corrected a critical bug in an otherwise competent translation. Medium-complexity benchmarks such as \texttt{httpie-xh} (+9.9 pp) and \texttt{jq-gojq} (+5.1 pp) show meaningful but more gradual improvement, suggesting the agent incrementally extends feature coverage across iterations.

However, iterative refinement has clear limits. Large benchmarks such as \texttt{taskwarrior} (+1.2 pp) and \texttt{charcoal-cli} ($-$1.0 pp) show marginal or negative change, indicating that when the initial implementation covers only a small fraction of required functionality, subsequent iterations cannot compensate for the fundamental scope gap. The average improvement of +11.0 pp suggests that iterative refinement is a viable strategy for benchmarks where the agent achieves a partial but buildable implementation in the first pass---precisely the regime where single critical bugs, rather than architectural gaps, are the dominant failure mode.

\end{document}